\documentclass{pasa}%

\usepackage{graphicx}
\usepackage{subfigure}
\usepackage{lscape} 

\title[91bg-like SNe host galaxies]{SN1991bg-like supernovae are associated with old stellar populations}

\author[F. H. Panther et al.]{Fiona H. Panther$^{1,2,3}$\thanks{E-mail: f.panther@adfa.edu.au (FHP)}, Ivo R. Seitenzahl$^{1, 2}$, Ashley J. Ruiter$^{1,2,3}$, Roland M. Crocker$^{2}$, Chris Lidman$^{2,3}$, Ella Xi Wang$^2$, Brad E. Tucker$^{2,4,5}$ and Brent Groves$^{2,4}$. 
\affil{$^1$School of Science, UNSW Canberra, Australian Defence Force Academy, Canberra 2612, Australia}
\affil{$^2$The Research School of Astronomy and Astrophysics, Mount Stromlo Observatory, Australian National University, Canberra, ACT 2611, Australia.}
\affil{$^3$ARC Centre of Excellence for All-Sky Astrophysics (CAASTRO)}
\affil{$^4$The ARC Centre of Excellence for All-Sky Astrophysics in 3 Dimensions (ASTRO 3D), Australia}
\affil{$^5$National Centre for the Public Awareness of Science, Australian National University, Canberra, ACT 2611, Australia.}
}%

\jid{PASA}
\doi{10.1017/pas.\the\year.xxx}
\jyear{\the\year}

\usepackage{aas_macros}
\usepackage{hyperref} 
\hypersetup{colorlinks,citecolor=blue,linkcolor=blue,urlcolor=blue}

\hypersetup{draft}

\begin{document}

\begin{frontmatter}
\maketitle

\begin{abstract}
SN1991bg-like supernovae are a distinct subclass of thermonuclear supernovae (SNe Ia). Their spectral and photometric peculiarities indicate their progenitors and explosion mechanism differ from `normal' SNe Ia. One method of determining information about supernova progenitors we cannot directly observe is to observe the stellar population adjacent to the apparent supernova explosion site to infer the distribution of stellar population ages and metallicities. We obtain integral field observations and analyse the spectra extracted from regions of projected radius $\sim\,\mathrm{kpc}$ about the apparent SN explosion site for 11 91bg-like SNe in both early- and late-type galaxies. We utilize full-spectrum spectral fitting to determine the ages and metallicities of the stellar population within the aperture. We find that the majority of the stellar populations that hosted 91bg-like supernovae have little recent star formation. The ages of the stellar populations suggest that that 91bg-like SN progenitors explode after delay times of $>6\,\mathrm{Gyr}$, much longer than the typical delay time of normal SNe Ia, which peaks at $\sim 1\,\mathrm{Gyr}$. 
\end{abstract}

\end{frontmatter}

\section{Introduction}
Type Ia supernovae (SNe Ia) are usually described as a photometrically and spectroscopically homogeneous class of astrophysical transients. They are thought to arise from the thermonuclear disruption of a carbon-oxygen (CO) white dwarf star in an interacting binary system (see \cite{Hillebrandtrev,Maguirerev} for a review). `Normal' SNe Ia are standardizable candles \citep{Branch92}: there is a tight relation between their peak luminosity and the width of their light curve (the \cite{Phillips93} relation), and between their peak luminosity and their optical color at peak luminosity \citep{Tripp98}. Thus they make an excellent tool for measuring cosmological distances. As standardizable candles, normal SNe Ia have been employed in cosmology to probe the geometry of the universe \citep{Riess98,Perlmutter99,Leibundgut01}. However, the proliferation of SN surveys over the past 30 years has led to the discovery of several spectroscopically and photometrically peculiar subclasses of these events. The discovery of SN1991bg in particular \citep{Fillipenko1992} - the prototype event for the SN1991bg-like subclass of SNe Ia (hereafter 91bg-like SNe, see \cite{Taubenberger17} for a review) -  challenged the accepted paradigm that SNe Ia are a homogeneous class of events.\\

91bg-like SNe events share several key features with their normal SNe Ia cousins, and are clearly thermonuclear supernovae. Specifically, they lack any indication of hydrogen and helium in their spectra while also exhibiting strong Si II in absorption \citep{Fillipenko1997}. Moreover, absorption features in their spectra near maximum light indicate the presence of a number of intermediate mass elements (IMEs) including silicon, magnesium, calcium, sulphur and oxygen. The presence of these IMEs is consistent with these events belonging to the class of thermonuclear transients \citep{Fillipenko1997}. However, 91bg-like SNe also exhibit significant photometric and spectroscopic peculiarities compared to normal SNe Ia. This is highly suggestive of a different explosion mechanism and/or progenitor configuration. The subclass differs from `normal' SNe Ia in the following ways: 
\begin{itemize}
	\item 91bg-like SNe are subluminous compared to SNe Ia by $1.5-2.5$ mag at optical maximum, reaching peak absolute magnitudes of only $-16.5$ to $-17.7$ in B \citep{Taubenberger17}, 
	\item 91bg-like SNe are significanly redder than their normal SNe Ia cousins, with $(B-V)_\mathrm{max} = 0.5-0.6$ mag, 
	\item 91bg-like SNe light curves decline very quickly, and are characterized by a lightcurve decline parameter $1.8\leq\Delta m_{15}(B)\leq2.1$, compared to $\Delta m_{15}(B)\leq1.7$ for normal SNe Ia. 91bg-like SNe also have a light curve rise time of $\sim13-15\,\mathrm{days}$ \citep{Taubenberger08}, a few days shorter than normal SNe Ia. Combined with their low peak luminosities, this is consistent with these events synthesising significanlty lower $^{56}$Ni masses than their normal SNe Ia counterparts ($\sim0.05 - 0.1\,M_\odot$, \cite{Sullivan11}). 
	\item 91bg-like SNe lack a secondary maximum in the near infrared (NIR), unlike normal SNe Ia, likely due to the 91bg-like SNe ejecta being cooler and less luminous than that of the normal SNe Ia. Instead, they exhibit a single NIR maximum delayed a few days with respect to the maximum in B \citep{Garnavich04}, whereas the first NIR peak in a normal SN Ia lightcurve preceeds the B band maximim. The absence of the secondary NIR maximuim makes it impossible to perform a simple time-shift to make the 91bg-like SNe lightcurve map to the standardizable lightcurve of the normal Ia, and the absence of the feature is physically interpreted as a merging of the primary and secondary NIR maxima into a single maximum due to the physical conditions in the SN ejecta \citep{Kasen06,Blondin15}.
	\item Spectroscopically, 91bg-like SNe exhibit similar pre-maximum optical spectra to normal SNe Ia - i.e. dominated by IMEs. However a transition to a spectrum dominated by iron group elements happens earlier in 91bg-like SNe than in normal SNe Ia \citep{Taubenberger17}.
	\item Particularly notable is the presence of unusually strong Ti II, and  O I$\lambda 7774$ in absorption in the post-maximum spectra. A number of these spectral peculiarities can be attributed to the unusually cool and slow-moving ejecta \citep[$\sim 7000\,\mathrm{km/s}$,][]{Taubenberger08} when 91bg-like SNe are compared to normal SNe Ia: the lower ionization state favours a higher abundance of neutral and singly ionized species. However, it is considered unlikely that ionization state and temperature explains all these spectral peculiarities.
\end{itemize}
To explain the spectroscopic and photometric peculiarities of 91bg-like SNe, a variety of progenitors and explosion mechanisms have been
suggested. A `violent merger' of two near-equal mass CO
WDs was explored via hydrodynamical simulations in \cite{Pakmor2010}.
Although the lightcurves of the these explosions were somewhat too 
broad to reproduce observations of 91bg-like SNe, the synthesised spectra, red colour and low expansion velocities were
a fairly good match to observed properties of 91bg-like SNe. To 
better reproduce the characteristics of 91bg-like SNe light curves,
it was suggested by \cite{Pakmor2013} that the
faint, fast lightcurves of 91bg-like SNe could result from the
merger of a CO WD and a helium WD (e.g. see their figure 4). 
The smaller ejecta mass that is expected in such an explosion can naturally explain the observed narrower lightcurves of 91bg-like SNe compared to their  `normal', brighter counterparts. 
A specific formation channel of the same 
progenitor scenario (CO WD + He WD merger) was found by \cite{Crocker17} to plausibly supply the high $^{44}$Ti yield that is necessary for these merging events to also 
explain the positron annihilation signal in the Milky Way.\\
While a number of progenitor configurations and explosion mechanisms have been suggested to explain the properties of 91bg-like SNe, no direct observations of the progenitor systems have been made, and so the exact progenitors of these SNe remain unknown. One method of inferring information about the progenitors of SNe independent of the observed properties of the explosion is to consider the galaxies in which they occur. A qualitative connection between host galaxy properties and the occurrence of 91bg-like SNe has already been well established. In LOSS (the Lick Observatory Supernova Search, \cite{Li2010}), around 15 per cent of all SNe Ia were of the 91bg-like SNe subclass. However, considering only early type galaxies, the fraction of SNe Ia that are of the 91bg-like SNe subtype rises to 30 per cent. Furthermore, in \cite{Perets10} the cumulative distribution of 91bg-like SNe occuring in galaxies of different Hubble types was found to be different to that of SNe Ia. While SNe Ia are observed to occur in all galaxy types, from passive to star forming, 91bg-like SNe occur predominantly in early type galaxies. While this is clear, a specific connection between the properties of early type galaxies (which tend to be more massive, more metal-rich, host significantly less star formation, and contain older stellar populations) and 91bg-like SNe is yet to be firmly established.\\
A number of works \citep{Gallagher08,Stanishev2012, Rigault2013, Galbany2014,Rigault2018} have now investigated the connection between the global properties of galaxies and the observed stretch and colour of SNe Ia, as this may be a source of bias in cosmology. While the focus has been on `normal' SNe Ia, \cite{Gallagher08} noted a correlation between older stellar populations and lower luminosity SNe Ia. Several of their low luminosity supernovae were classified spectroscopically as SNe 91bg in the Berkeley Supernova Project \citep{Silverman12}, and their host galaxies were found to contain stars with an average age of $\sim 10\,\mathrm{Gyr}$.\\
The observations of \cite{Gallagher08} were restricted to early-type galaxies and utilized slit spectroscopy. Consequently, the ages and metallicties of the stellar populations derived represent a single global average for these SNe host galaxies. Since these host galaxies are early-type galaxies where the stellar populations will be well-mixed, the global galaxy properties are a good approximation for the stellar population that gives rise to the supernova progenitor \citep{Galbany2014}. However, the method employed by \cite{Gallagher08} makes it hard to exclude the presence of recent local star formation within a few kpc of the supernova progenitor. Furthermore, the \cite{Gallagher08} work makes use of the Lick/IDS Index system \citep{Worthey94, Worthey97} to derive the average age and metallicity of the stellar populations of the observed galaxies, and does not take into account that stellar population will be composed of stars that formed at different times and thus cannot be characterized by a single stellar age.\\
Global galaxy properties may help with the standardization of supernova lightcurves, however to determine whether a supernova subtype is associated with an old ($>2-3\,\mathrm{Gyr}$ old) stellar population requires one to rule out the presence of recent star formation within a few kpc of the apparent supernova explosion site.\\
To do this, it is necessary to consider a spectrum of the stellar population adjaced to the apparent supernova explosion site. Integral Field Unit (IFU) observations can be utilized to investigate this. For nearby supernova hosts (up to $z\sim 0.03$), an aperture with a radius of $\sim 1.5"$ can resolve a region around $1\,\mathrm{kpc}$ in projected radius on the galaxy. Furthermore, a local measurement can also be used to determine the distribution of stellar population ages using a method such as penalized pixel fitting of the stellar spectrum extracted from the same apeture used in step one \citep{ppxf1, ppxf2}.\\
In this work, we will utilize penalized pixel fitting to investigate the age distribution of the stellar populations that hosted 91bg-like supernovae in a range of galaxies of all morphophological types. We choose to use the local stellar spectra at the supernova explosion site as opposed to integrated galaxy spectra. For example, in the case of a late-type host galaxy, star formation may be localized to the disk of a galaxy. In an integrated spectrum of the galaxy, this star formation can dominate the observed light due to the luminosity of O-A stars. Thus, if a supernova occurs in the bulge of a spiral galaxy, one is likely to underestimate the age of the population that gave rise to the SN progenitor of only global host galaxy properties are considered. On the other hand, if a SN occurs in an early-type galaxy where significant stellar migration may have occured, one can still use the IFU technique to constrain the age of the supernova progenitor: if a SN progenitor is formed within the last Gyr, evidence of this star formation will still be apparent in the stellar spectra as stars will not migrate far from their birthplaces in this timeframe \citep{Fujii2012}. Furthermore, the local stellar population will have a distribution of ages representative of that of the whole galaxy \citep{Galbany2014}.\\ 
Several works have utilized a similar method, employing IFU spectroscopy to investigate the region in the immediate projected vicinity of SNe Ia to determine the relationship between the scatter in SNe Ia properties after standardization, and star formation rate \citep{Stanishev2012, Rigault2013, Rigault2018}. Those works are particularly important in the context of cosmology, as the dispersion in Hubble residuals can affect the precision and accuracy of measurement of the cosmological parameters, such as the Hubble constant and the dark energy equation of state parameter. However, the properties of stellar populations at the explosion sites of different subtypes of SNe Ia have not yet been studied in detail. 91bg-like SNe are a distinct subclass which are empirically shown to occur predominantly in old stellar populations. Thus, they provide a perfect opportunity to use IFU spectroscopy to quantify the age of stellar populations close to the SN explosion site and constrain the presence of recent star formation.\\
In this paper, we present the results of our observations of 91bg-like SNe host galaxies using IFU spectroscopy based on data obtained at Siding Spring Observatory. We derive the distribution of stellar population ages and metallicities of the stars in a region of radius $\sim 1\,\mathrm{kpc}$ around the apparent SN explosion site using full-spectrum fitting. Our observations and data reduction are described in section \ref{sec:obs}, and the results of the full-spectrum fitting procedure is described in section \ref{sec:results}. Finally, we present a discussion of our results in section \ref{sec:discussion} and our conclusions in section \ref{sec:conclusion}.
\section{Observations and Data Reduction} \label{sec:obs}
\subsection{Sample selection}
\begin{table*}
	\centering
	\caption{Summary of observations.}
	\label{tab:obs}
	\begin{tabular}{lllrccccc}
		\hline
		SN Name & Host Galaxy & SN RA & SN DEC & t$_\mathrm{exp}$ (s) & Morph$^1$ & A$_\mathrm{ap}$ ($\mathrm{kpc^2}$) & z\\
		\hline
		SN1993aa & Anon J230322-0620 & 23 03 22.02 & -06 20 56.1 & 2$\times$1800 & 10 & 1.23&0.018\\
		SN2000ej & IC1371 & 21 20 15.66 &  -04 52 40.5 & 2$\times$1800 &  -5 & 3.6&0.0308\\
		SN2007ba &	UGC9798	& 15 16 41.83	&  +07 23 48.1&2$\times$1800 &  0 & 5.5&0.03825\\
		SN2007cf &	MCG+02-39-021 &	15 23 07.66	  & +08 31 45.5	&	2$\times$1800 &  -2 & 4.1&0.03293\\
		SN2007fq &	MCG-04-48-19 &	20 34 55.92	& -23 06 15.84 &	2$\times$1800	& -2 & 6.8&0.04245\\
		SN2012fx$^{2}$ &	ESO417-03 & 02 55 41.20	& -27 25 27.59	& 4$\times$1800 &  5 & 1.2&0.0176\\
		SN2002cf &	NGC4786	& 12 54 31.30	& -06 51 24.8	& 3$\times$1200 & -5 & 0.9&0.015\\
		SN2002jm &	IC603	& 10 19 25.20	& -05 39 14.5	& 2$\times$1200	&  1 & 1.3&0.0182\\
		SN2007al	& Anon J095919-1928 & 	09 59 18.48	& -19 28 25.8 &	3$\times$1200	&-5 & 0.6&0.0121\\
		SN2008ca &	SDSS J122901.28-263305.6	& 12 29 01.20 &	 +12 22 18.84	& 3$\times$1200	 & 10 & 55&0.123\\
		SN2008bt &	NGC3404	& 10 50 16.99	& -12 06 31.5	& 3$\times$1200s& 	1 & 0.91&0.0154\\
		\hline
		SN2002ey$^4$ & Anon J23101188+0732541	& 23 10 12.33 & +07 32 59.9 & 2$\times$1800 &  -1 & 5.7&0.0388\\
		SN2005er$^4$	& NGC7385	& 22 50 00.84	&  +11 37 05.7&	3$\times$1800 &	-2 & 2.6&0.02616\\
		SN2005ke$^4$ &	NGC1317	& 03 35 04.35	&  -24 56 38.8&	3$\times$1200 &  0 & 0.2&0.00643\\
		SN2007N$^4$	& MCG-01-33-12 &	12 49 01.25	& -09 27 10.2	& 3$\times$1200	&  1 & 0.6&0.0126\\
		SN2006gt$^4$ &	Anon J005618-0137 &	00 56 17.30	& -01 37 46.0 &	3$\times$1800	& -5 & 7.5&0.0447\\
		lsq15bb$^{3,4}$&	 2dFGRS TGN352Z120	& 09 58 51.95	& +01 01 04.12&	3$\times$1200 & 	0 &  23&0.08\\
		\hline
	\end{tabular}\\
    \begin{flushleft}
	$^1$ - Morphology based on DSS R-band images classified by the authors using numerical Hubble types where negative numbers represent early-type galaxies, and positive numbers late-type \citep{devac}.\\
	$^2$ - SN classification from \cite{2012fx}.\\
	$^3$ - SN classification from \cite{lsq15bb}.\\
    $^4$ - Removed from sample due to low SNR in datacube due to adverse weather conditions or SN at large offset from host.\\
    \end{flushleft}
\end{table*}
 Our sample of 91bg-like SNe host galaxies is composed of 15 hosts of SNe spectroscopically identified as 91bg-like SNe by \cite{Silverman12} visible from Siding Spring Observatory, as well as the host of SN2012fx \citep{2012fx} and the host of lsq15bb \citep{lsq15bb}. We do not make any selection cuts based on galaxy morphology, luminosity or mass. Nor do we make a selection cut against supernovae that have large offsets from their host as there may be a diffuse stellar population present at the explosion site not obvious in the Digitized Sky Survey 2 (DSS-2\footnote{http://archive.eso.org/dss/dss}) images which were used to create finder charts for each observation. 
\subsection{Observations}
We obtained integral field observations of 17 spectroscopically identified 91bg-like SNe host stellar populations using the WiFeS instrument \citep{Dopita07} on the Australian National University 2.3m Advanced Technology Telescope at Siding Spring Observatory (SSO).\\
Observations were obtained over a total of 14 nights between July 2016 and February 2017. WiFeS is a wide-field Integral Field Unit with a total field of view of $25"\times38"$. The instrument configuration is shown in Table \ref{tab:setup}. The aperture of WiFeS is centered on the apparent location of each SN, with the position angle of the aperture chosen to minimize contamination from nearby stars but, as we only extract spectra from pixels close to the apparent SN location, the position angle has little practical importance.\\
Exposure times reflect changes in our observation strategy to account for inclement weather conditions. A summary of all our science observations is given in Table \ref{tab:obs}. Bias frames, Cu-Ar arcs and dome flats were obtained at the beginning and end of the night. Wire frames\footnote{A wire frame ensures the spectrograph slits are appropriately aligned on the CCD} were obtained at the end of each night, and sky flats were obtained at the beginning of each night. For each individual galaxy observation, Cu-Ar arcs (required for accurate wavelength solution, which may drift a few Angstrom over the course of a night) and bias frames are obtained within 30 minutes of the observation, and each galaxy observation is accompanied by at least one sky frame, which is used to remove skyline contamination. Finally, a spectrophotometric standard star is observed on each night of observation. Where weather conditions did not permit time for these observations, spectrophotometric calibration was carried out using a spectrophotometric standard taken in earlier runs, with the standard frames being taken in similar photometric weather conditions and at similar airmass to the galaxy being observed.\\
\begin{table}
	\centering
	\caption{WiFeS instrument Configuration}
	\label{tab:setup}
	\begin{tabular}{ll}
		\hline
		Component & Configuration\\
		\hline
		Red Grating & R3000\\
		Blue Grating & B3000\\
		Dichroic & RT560\\
		Wavelength range & $\sim3300$ \AA - $\sim9200$ \AA \\
		\hline
	\end{tabular}
\end{table}
 \subsection{Data Reduction}
 Data reduction was facilitated by the \texttt{PyWiFeS} data reduction package (\cite{PyWiFeS}). The pipeline produces a spectrophotometrically calibrated and fully coadded datacube, and a corresponding parallel-processed error cube. At this point, a number of SN hosts are removed from the sample due to the observations having either been undertaken in inclement weather (footnote 4 of \ref{tab:obs}), or of a SN explosion site with a large offset from the SN host galaxy, thus the SNR in the datacube is inadequate to obtain either a local or global stellar spectrum associated with the SN host\\
Spectra of the stellar populations local to each SN were then extracted from the calibrated datacube. We use a seeing-limited aperture of $3"\times3"$ centered on the apparent location of the SN in each host galaxy (the physical size of these apertures varies with redshift and is shown in Table \ref{tab:obs}). These spectra are processed to remove residual sky lines using a nearby area free from galaxy light, offset from the host galaxy from the same final datacube, and the mean resulting spectrum is extracted from the $3"\times3"$ region of interest.\\
Each remaining spectrum is then shifted to rest-frame wavelength according to the host redshift listed on NED\footnote{https://ned.ipac.caltech.edu/ The NASA/IPAC Extragalactic Database (NED) is operated by the Jet Propulsion Laboratory, California Institute of Technology, under contract with the National Aeronautics and Space Administration.}, which is checked against the redshift we derive from either the narrow H$_\alpha$ line if present, or the H and K lines for the remaining galaxies. The spectra are then corrected for Milky Way dust reddening using \texttt{brutus 0.3.1}\footnote{http://fpavogt.github.io/brutus/index.html}. Galactic extinction is derived using NASA Extragalactic Database (NED) from the \cite{Schlafly11} recalibration of \cite{Schlegel98} infrared based dust map. The map is based on dust emission from COBE/DIRBE and IRAS/ISSA; the recalibration assumes a \cite{Fitzpatrick1999} reddening law with $Rv = 3.1$ and different source spectrum than \cite{Schlegel98}. The resulting spectra are subsequently ready for analysis.
\section{Results}\label{sec:results}
\subsection{Full Spectrum Fitting}
Deriving stellar population ages requires simultaneous determination of both the age of stars and their metallicity as there is a significant degeneracy between the effect of metallicity and age on galaxy spectra \citep{Worthey94}.\\
\begin{figure*}
\centering
\subfigure[Stellar pop. within 0.57 kpc of SN1993aa]{\label{fig:a}\includegraphics[width=80mm]{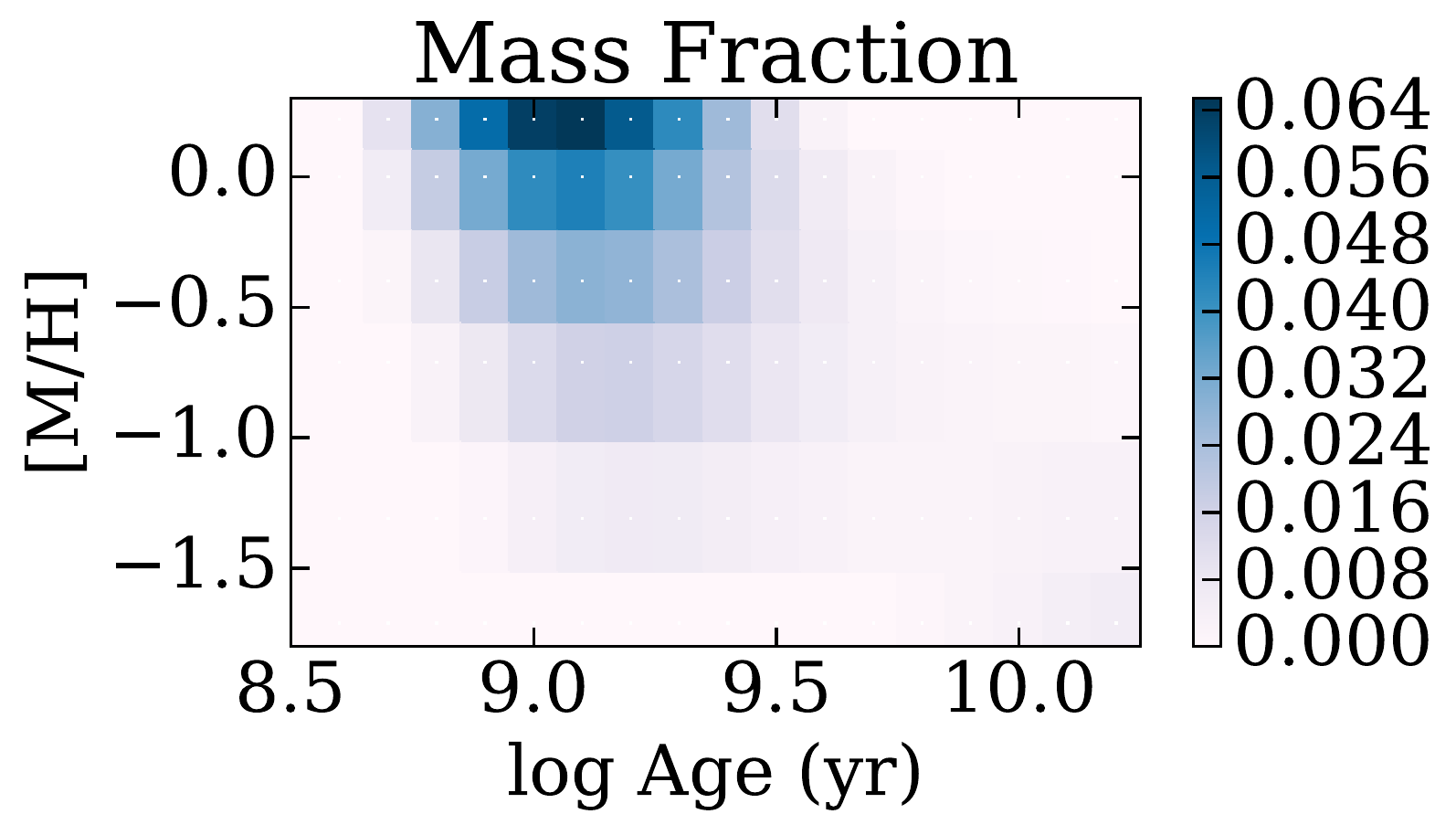}}
\subfigure[Stellar pop. within 0.96 kpc of SN2000ej]{\label{fig:b}\includegraphics[width=80mm]{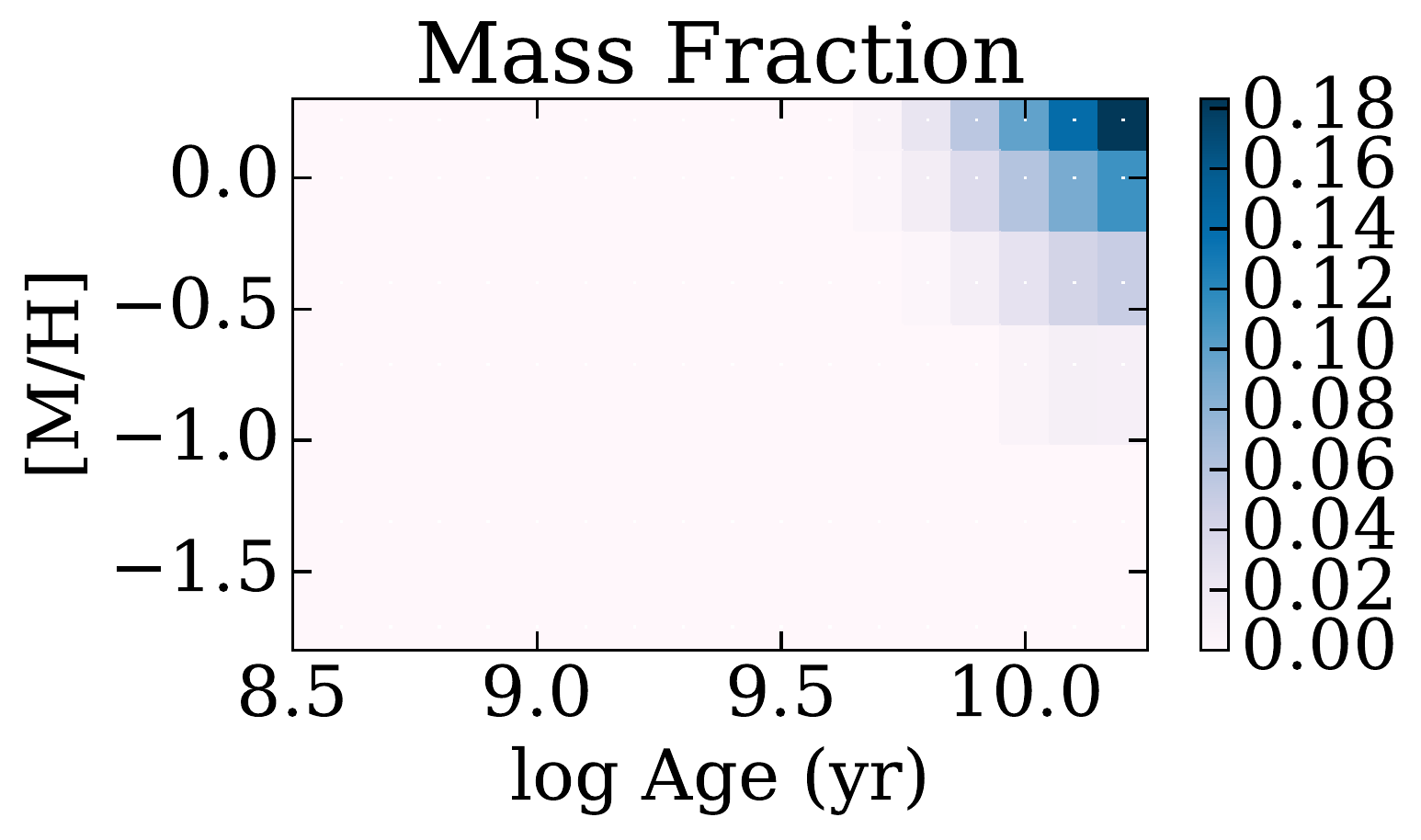}}
\subfigure[Stellar pop. within 0.49 kpc of SN2008bt]{\label{fig:i}\includegraphics[width=80mm]{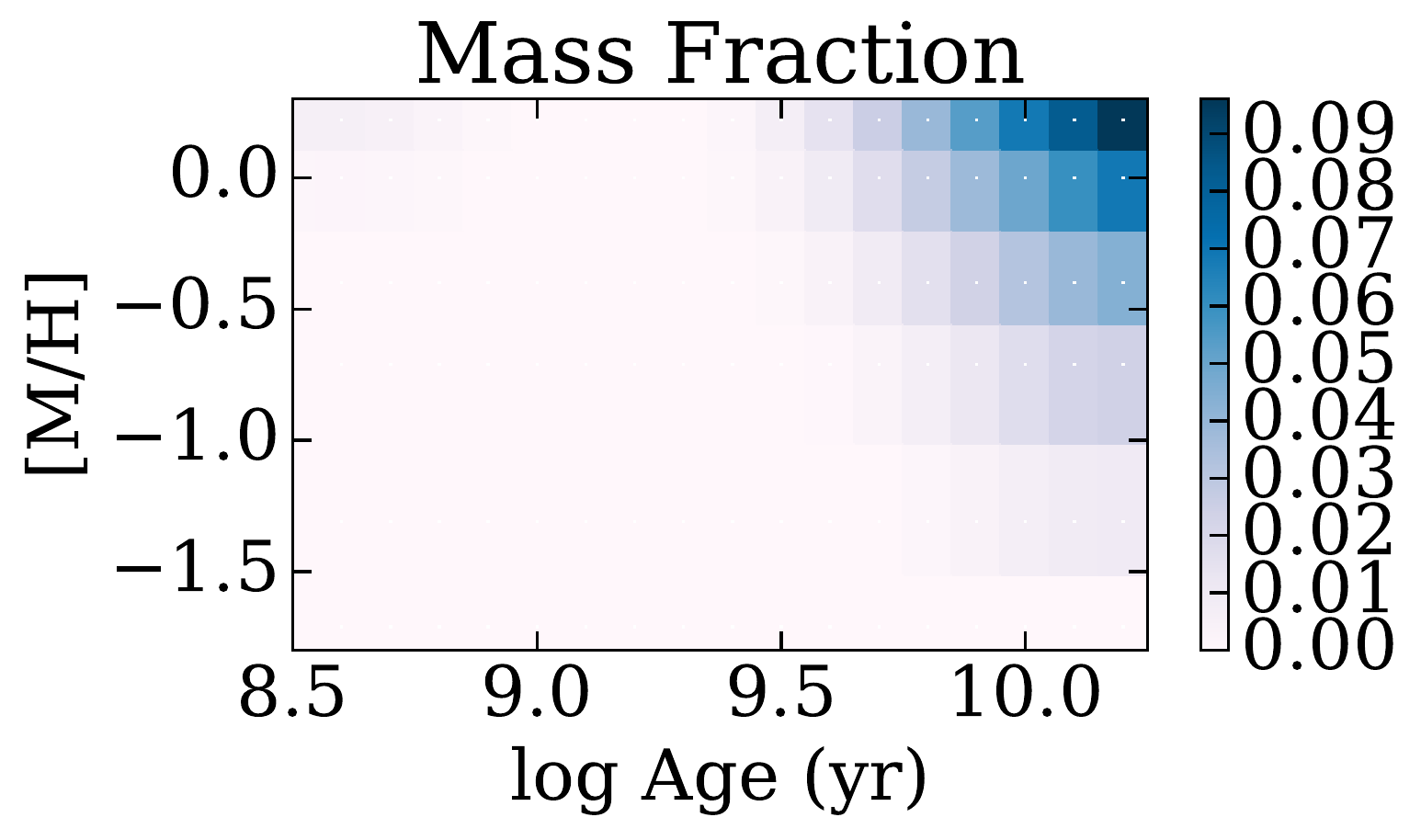}}
\subfigure[Stellar pop. within 3.44 kpc of SN2008ca]{\label{fig:j}\includegraphics[width=80mm]{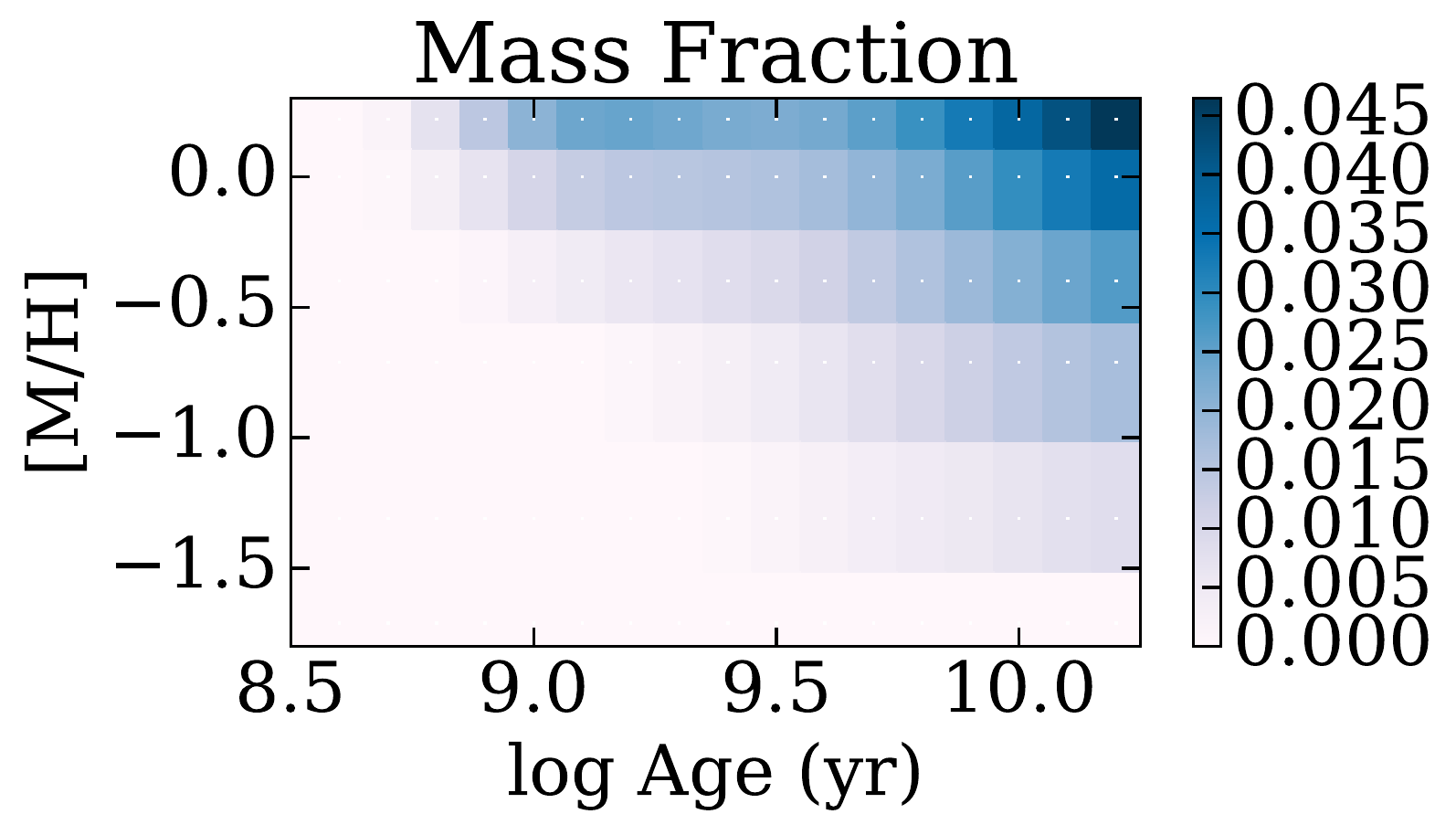}}
\caption{\label{fig:weightedbins1}Mass fraction of stars in different age and metallicity bins based on spectra extracted from 3"x3" regions centered on the SN explosion site in each galaxy, which are interpreted as representing the distribution of stellar population ages which give rise to the SN progenitor. The radius of the aperture in physical units is shown in each subcaption. The population shown in (b) is representative of the remaining seven galaxies for which the stellar populations are fitted with \texttt{pPXF}.}
\end{figure*}
We use the publicly available penalized pixel fitting code (\texttt{pPXF}, \cite{ppxf1,ppxf2}) to fit linear combinations of simple stellar populations (SSPs) from the MILES stellar library \citep{MILES,vazdekis2010}, which provides $985$ stellar spectra across a wavelength range of $3525-7500$\,\AA\,at 2.5\,\AA\,spectral resolution, for stellar metallicities ranging from $[M/H] = 0.22$ to $[M/H] = -2.3$. The `solution' of the problem - to fit multiple SSPs to an observed spectrum - is defined by different weights applied to each SSP in a regular grid in log-age and metallicity space. The program outputs a weighted average age of the stellar population and a weighted average metallicity according to these weights, along with information about the weights applied to each age-metallicity bin. This allows us to determine both the distribution of stellar population ages that contribute to the spectrum, and the weighted average age of the stellar population, as well as whether any substantially younger or older stellar population than the average is present.\\
We obtain a fit to the total spectrum of 11 91bg-like SN host galaxies (with flux normalized by the flux at $5500$ \AA). \texttt{pPXF} decomposes the spectra into nebular and stellar components, and determines the mass fraction of stars into age and metallicity bins determined by the available grid of SSP models.\\
In Fig. \ref{fig:weightedbins1} we show outputs from \texttt{pPXF} which display the mass fraction of stars in age and metallicity bins. The majority of stellar populations we observe exhibit similar characteristics to Fig. \ref{fig:b}, including the local stellar population of SN2012fx, which occured in the bulge of a late-type galaxy. These stellar populations have a large mass fraction of stars at ages $\sim 10\,\mathrm{Gyr}$ at super-solar metallicities ($[M/H]>0$). The weighted average stellar population properties of seven of our observed SN explosion sites suggest that 91bg-like SNe occur preferentially in stellar populations with ages $\sim 10\,\mathrm{Gyr}$ with solar, or slightly sub-solar metallicities (Fig \ref{fig:map}).\\
Three of the observed stellar populations, however, contain a younger stellar component: the host of SN1993aa (Fig \ref{fig:a}), the host of SN2008bt (Fig \ref{fig:i}) and the host of SN2008ca (Fig \ref{fig:j}). The spectra of these galaxies are also unusual in comparison to the majority of the galaxies observed. An example of a typical spectrum we analyse with \texttt{pPXF} is shown in Fig. \ref{fig:baspec}. This spectrum corresponds to the population shown in Fig \ref{fig:b}, and is composed predominantly of old, metal rich stars. In comparison, the spectrum of the host population of SN1993aa (Fig \ref{fig:aaspec}) is significantly bluer and exhibits strong emission lines of H$_\alpha$, [N II], [S II], [O III]\,and H$_\beta$. The strong emission lines indicate there may be ongoing star formation in this stellar population, in good agreement with the high mass fraction of stars with ages $\sim 1\,\mathrm{Gyr}$ seen in Fig \ref{fig:a}.  Similarly, the host population of SN2008ca also exhibits strong emission lines (Fig \ref{fig:caspec}), however the spectrum is (within the noise) somewhat less blue than that of SN1993aa, suggesting an older stellar component may also be present (Fig \ref{fig:j}). While the physical area of the aperture within the host of SN1993aa encompasses $\sim 1\,\mathrm{kpc}$, it should be noted that the host of SN2008ca is significantly more distant than that of SN1993aa and therefore the entire galaxy, with an area of $\sim 50 \,\mathrm{kpc}$ is sampled in the aperture. This may be the reason we observe apparent multiple stellar populations in this SN host galaxy. An example of a more resolved galaxy that also exhibits multiple stellar populations is the host of SN2008bt. This galaxy has strong H$_\alpha$, [N II] and [S II] emission (fig \ref{fig:btspec}), and while it has a predominantly old stellar population, there is a small mass fraction of stars which have an age of $\sim 0.5\,\mathrm{Gyr}$ (fig \ref{fig:i}). When we consider the position of the aperture on the host galaxy, we see that while the SN has occurred in a position that may be coincident with the bulge of the galaxy, because the galaxy is viewed almost edge on, there may be some contamination of starlight and nebular emission from a younger, star-forming disk. While we cannot rule out that the supernova occurred in this region, we do find that the majority of the stellar mass within the aperture is concentrated in the old, metal rich stellar population.
\begin{figure*}
\centering
\subfigure[Fitted spectrum of SN2000ej host]{\label{fig:baspec}\includegraphics[width=0.4\textwidth]{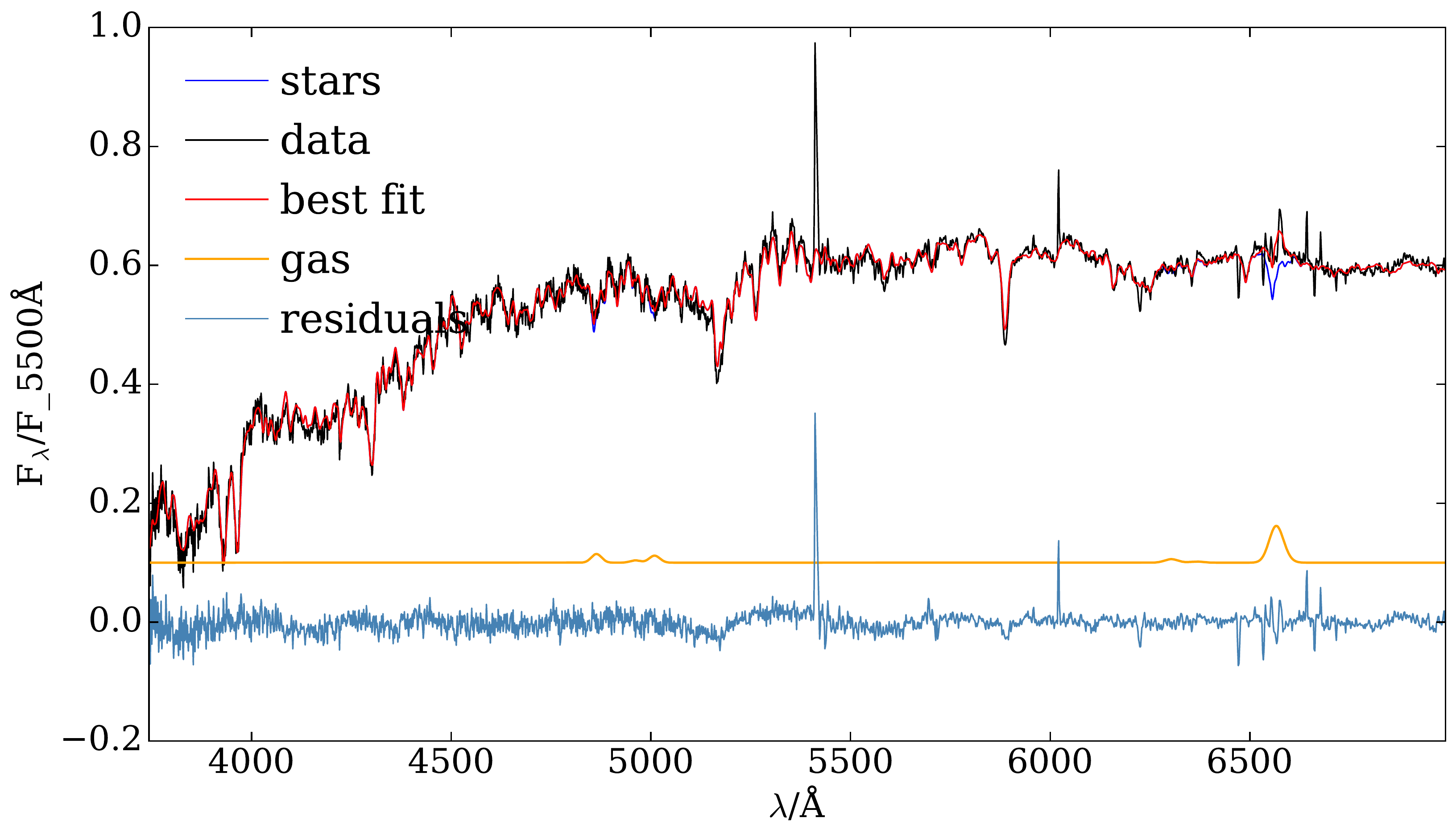}}
\subfigure[Fitted spectrum of SN1993aa host]{\label{fig:aaspec}\includegraphics[width=0.4\textwidth]{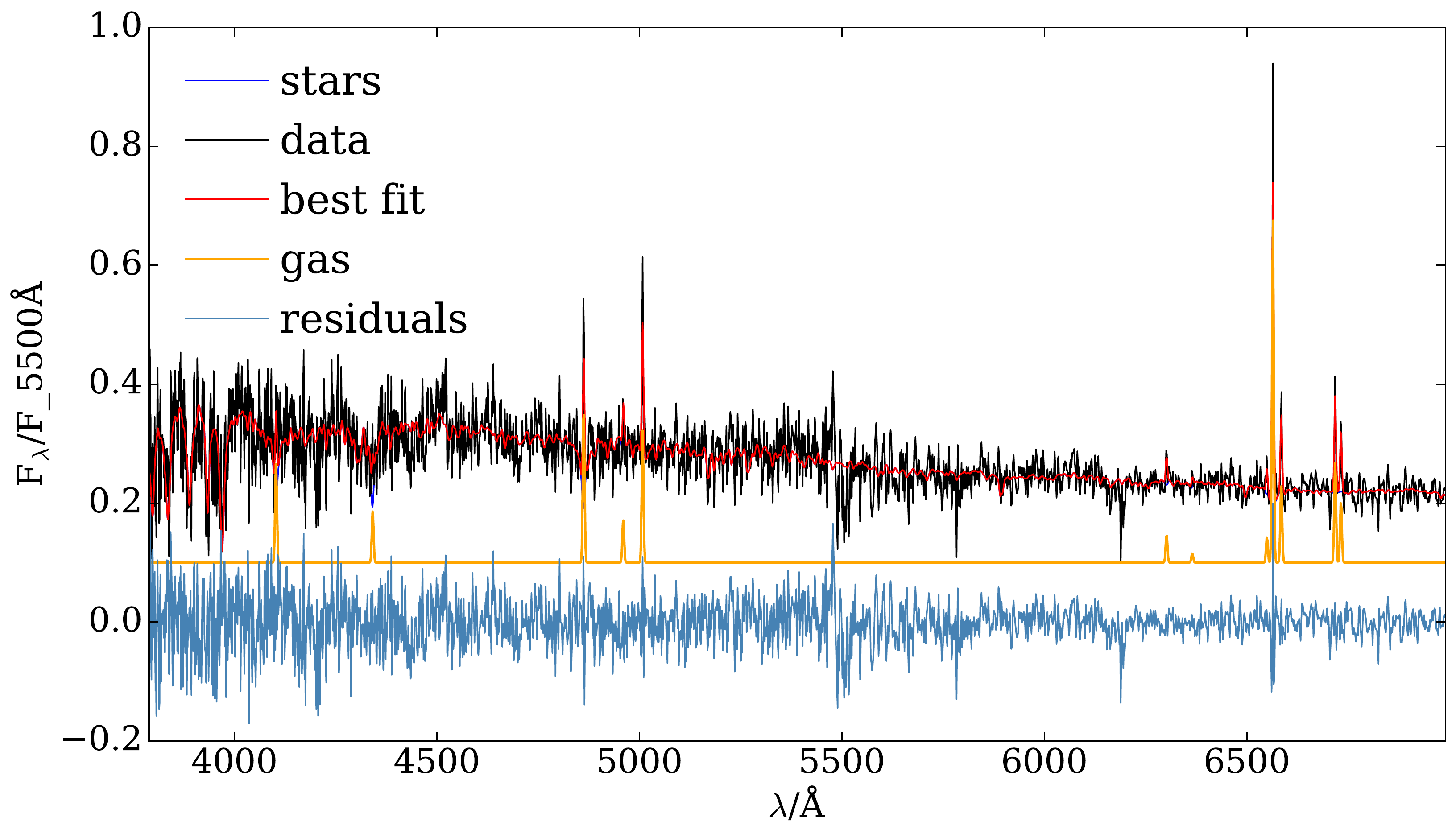}}
\subfigure[Fitted spectrum of SN2008bt host]{\label{fig:btspec}\includegraphics[width=0.4\textwidth]{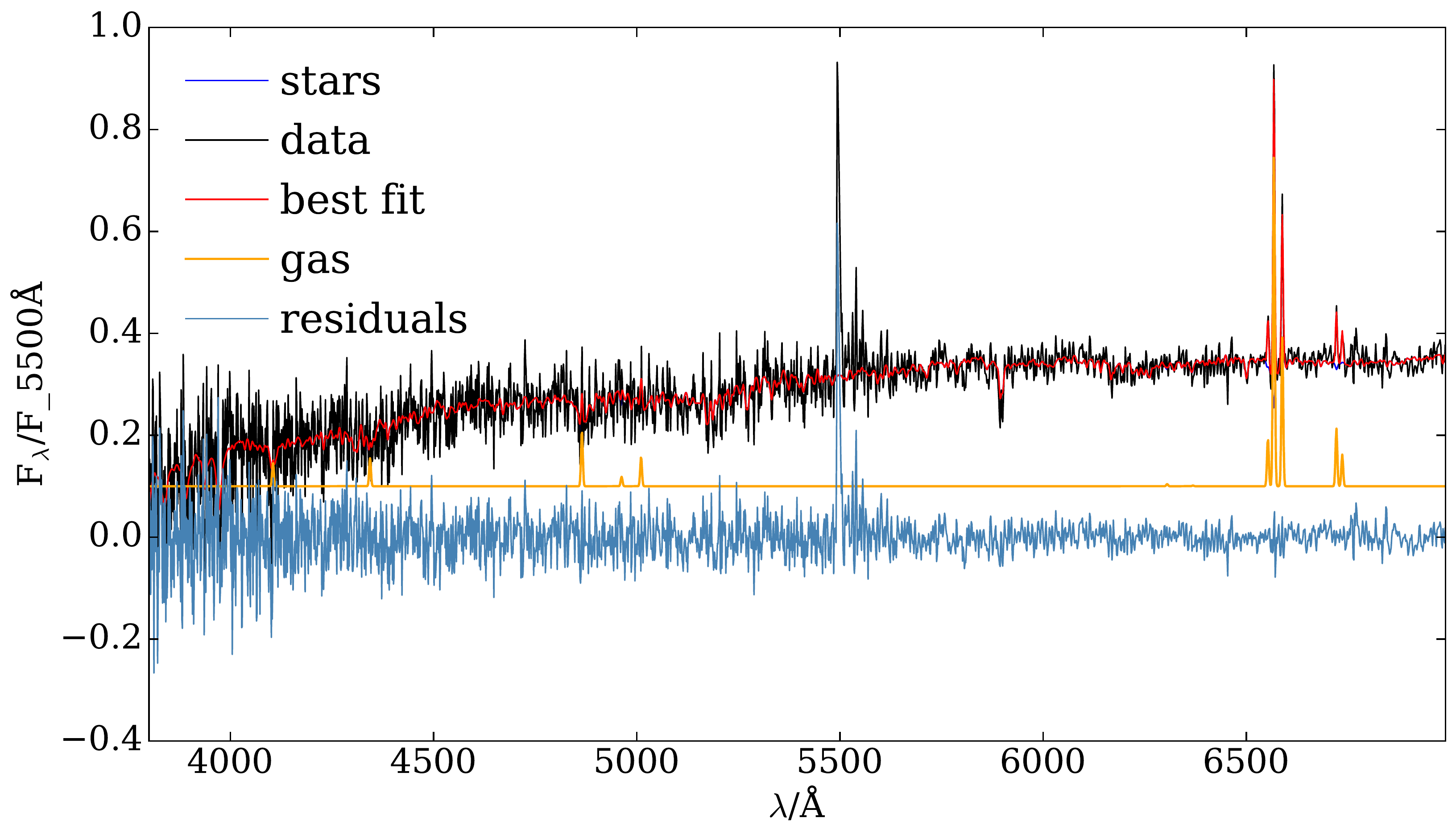}}
\subfigure[Fitted spectrum of SN2008ca host]{\label{fig:caspec}\includegraphics[width=0.4\textwidth]{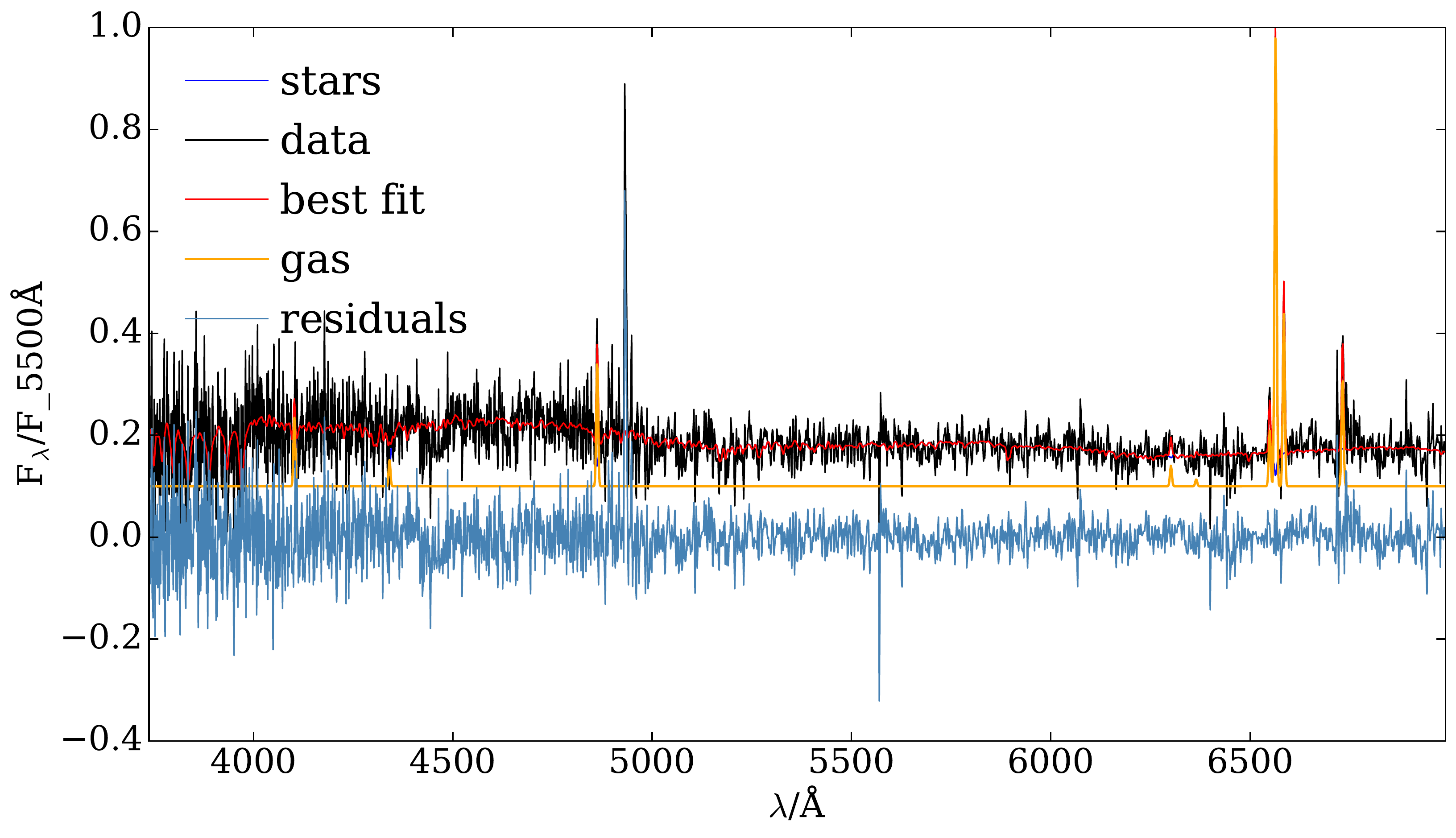}}
\caption{\label{fig:spectra}Spectra fitted using \texttt{pPXF}. Data is shown in grey, and the best fit to the total spectrum (nebular plus stars) is shown in red. The spectrum is also decomposed into stellar(dark blue) and nebular spectra (orange). Residuals for the best fit are shown in light blue. For several of the spectra, there are substantially larger residuals toward bluer wavelengths due to the lower SNR at blue wavelengths.}
\end{figure*}
\subsection{Comparison sample of SNe Ia}
While we wish to quantify the ages and metallicities of the stellar populations that host 91bg-like SNe, we also wish to compare the distribution of stellar population ages and metallicities to a sample of SNe Ia host galaxies to investigate whether 91bg-like SNe differ in some way. Unfortunately, a representative sample of SNe Ia host galaxies with spectra of the stellar population in the immediate vicinity of the apparent SN location is not publicly available. However, in \cite{Maoz2012} a sample of SDSS galaxies that hosted SNe Ia is compiled. As a comparison sample, we compute the weighted average stellar population age and metallicity for these galaxies using \texttt{pPXF}. We note that this average is for the integrated light of the host galaxy where the SDSS fiber is centered on the core of the galaxy, and not solely the stellar population $\sim 1\,\mathrm{kpc}$ from the apparent supernova location, however it has been previously noted that the global properties of SNe Ia host galaxies correlate well with the properties local to the SN explosion site \citep{Galbany2014}.
\begin{figure*}
	 		\includegraphics[scale=0.5]{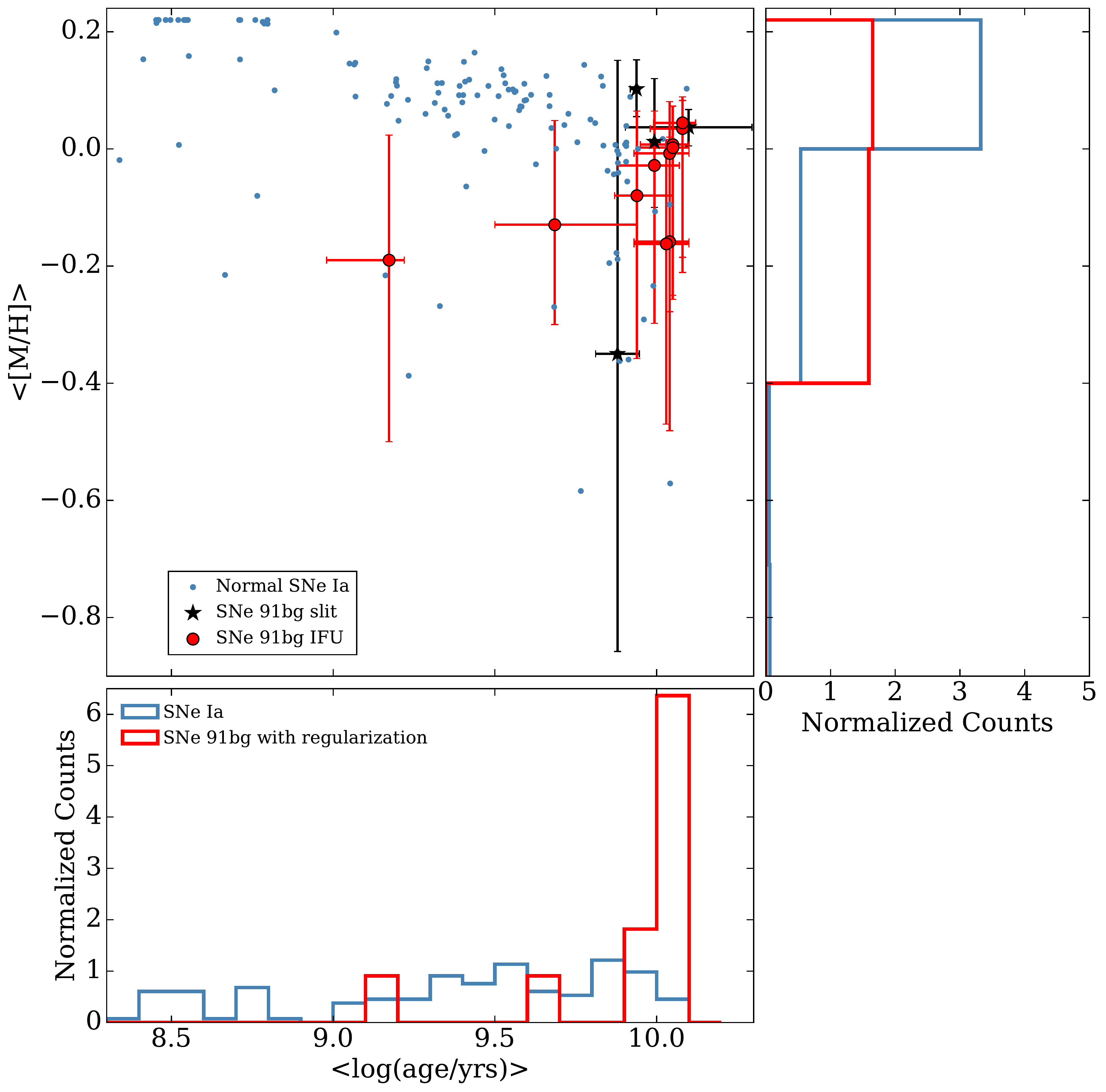}
	 		\caption{\label{fig:map} Weighted average stellar population ages for SDSS spectra of galaxies that hosted SNe Ia, based on the sample presented in Maoz et al. 2012 (blue) and for the local stellar populations of SNe 91bg galaxies (red). Error bars show the 16th and 84th percentile of the distrubution of stellar population ages and metallicities based on the CDF of stellar population properties. Black points show the ages and metallicities of SNe 91bg host galaxies observed in Gallagher et al. 2008, derived using Lick/IDS indicies and their respective errors. The histograms show the distribution of weighted average stellar population ages (bottom) and metallicities (right). The binning reflects the available metallicities and ages of the stellar population templates in the MILES library used to fit the spectra.}
\end{figure*}

\begin{table*}
	\centering
	\caption{Summary of results}
	\label{tab:res}
	\begin{tabular}{llcc}
		\hline
		SN Name & Host Galaxy &$<\log(\mathrm{age/yrs})>$ & $<[M/H]>$\\
		\hline
		SN1993aa & Anon J230322-0620 & $9.173_{-0.193}^{+0.047}$ & $-0.19_{-0.31}^{+0.21}$\\
		SN2000ej & IC1371 &  $10.08_{-0.19}^{+0.05}$ &$0.03_{-0.25}^{+0.05}$\\
		SN2007ba &	UGC9798	& $10.04_{-0.11}^{+0.06}$&  $ -0.01_{-0.27}^{+0.09}$\\
		SN2007cf &	MCG+02-39-021 &	$10.05_{-0.11}^{+0.05}$	  & $0.00_{-0.26}^{+0.07}$\\
		SN2007fq &	MCG-04-48-19 &	$10.08_{-0.09}^{+0.04}$	& $0.04_{-0.23}^{+0.04}$\\
		SN2012fx$^{2}$ &	ESO417-03 & $9.993_{-0.11}^{+0.08}$	& $-0.03_{-0.27}^{+0.09}$\\
		SN2002cf &	NGC4786	& $10.04_{-0.11}^{+0.06}$	& $-0.16_{-0.32}^{+0.19}$\\
		SN2002jm &	IC603	& $10.05_{-0.10}^{+0.05}$	& $0.01_{-0.26}^{+0.07}$\\
		SN2007al	& Anon J095919-1928 &$10.03_{-0.10}^{+0.07}$& $ -0.16_{-0.31}^{+0.18}$\\
		SN2008ca &	SDSS J122901.28-263305.6& $9.685_{-0.185}^{+0.255}$&	 $-0.13_{-0.17}^{+0.18}$\\
		SN2008bt &	NGC3404	& $9.939_{-0.07}^{+0.11}$	& $-0.08_{-0.28}^{+0.14}$\\
		\hline
	\end{tabular}\\
\end{table*}

We find that the weighted average ages of stellar populations have a wide, almost uniform distribution, with ages between $\sim 30\,\mathrm{Myr}$ and $>10\,\mathrm{Gyr}$. The metallicity distribution is peaked toward super-solar metallicities ($[M/H]>0$). We also find that the older the stellar population of the galaxy, the more metal-poor the weighted average metallicity becomes.      
\subsection{Stellar populations hosting SNe 91bg}
As a further comparison, we overplot the stellar population ages and metallicities determined using slit spectroscopy for early type galaxies hosting SNe 91bg from \cite{Gallagher08}. The black markers in Figure \ref{fig:map} show the ages and metallicities of these galaxies. These ages and metallicities are determined using Lick indicies \citep{Worthey94,Worthey97} which relies on comparing the measured equivalent width of different absorption lines found in galaxy spectra to an irregular diagnostic grid computed from simple stellar populations. As these galaxies were observed using slit spectroscopy, the stellar populations are representative of the average age of the stars contained in the whole galaxy. Compared to the SNe Ia comparison sample taken from \cite{Maoz2012}, the galaxies that host SNe 91bg in \cite{Gallagher08} appear to contain older stellar populations. Furthermore, the ages and metallicities derived for these galaxies are consistent with the ages and metalliticies of the local stellar populations of SNe 91bg we derive from our observations.\\
\section{Discussion}\label{sec:discussion}
In this work we analyzed the spectra of galaxies observed with the WiFeS IFU. This allowed us to isolate the stellar population within a few kiloparsecs of the apparent SN explosion site, and determine the age and metallicity of the stellar population in the vicinity of the SN. The age of the stellar population surrounding the SN explosion site in particular may constrain the delay time distribution of 91bg-like SNe (the distribution of times between the SN progenitor formation and the subsequent SN explosion).\\
In particular, we find that the stellar populations that host SNe 91bg have greater weighted average stellar population ages than those of SNe Ia and many lack evidence of recent star formation. This is strong evidence that 91bg-like SNe are not associated with recent star formation (within the last $\sim 1\,\mathrm{Gyr}$, as the stellar spectra lack evidence of a significant number of O-A stars being present. Furthermore, these results suggest that the delay time distribution of 91bg-like SNe is peaked toward very long delay times ($10\,\mathrm{Gyr}$), in contrast with SNe Ia, whose delay times peak at $\sim1\,\mathrm{Gyr}$ \citep{Childress2014}. We compared our derived average stellar population ages and metallicities for the local stellar populations of SNe 91bg with the stellar population ages and metallicities of SNe 91bg host galaxies found by \cite{Gallagher08} and find a good agreement within the error bars between the two results.\\
We note, however, the large errors on the metallicity of the stellar populations in both our own measurements and those from \cite{Gallagher08}. While two different methods are used to determine the stellar population parameters - full spectrum fitting with \texttt{pPXF} in this work and the Lick/IDS index system  in \cite{Gallagher08} - the stellar model library used to compute stellar population ages and metallicities is identical \citep[the MILES library][]{vazdekis2010}. While these stellar templates are well sampled in the age parameter space, there are only seven available metallicities between $[M/H] = -2.3$ and $[M/H] = 0.2$. Although the metallicity computed by full spectrum fitting is accurate, it lacks the precision that would be possible with a finer grid of models in metallicity space. This is reflected by the error bars shown in Fig \ref{fig:map}.\\
The SNe 91bg host galaxies from \cite{Gallagher08} are part of a larger sample of massive elliptical galaxies that hosted thermonuclear supernovae. As massive elliptical galaxies typically contain older stellar populations, there is likely a bias toward sampling host stellar populations where we expect the relative rate of SNe 91bg to normal SNe Ia to be somewhat higher than average. In our sample, we make no selection cuts and select galaxies of any morphological classification that hosted SNe 91bg.
The majority of the host galaxies we observe are selected from the BSNIP sample \citep{Silverman12}. These supernovae were observed as part of a targeted, local survey. Consequently, the galaxies in which SNe in a targeted survey occur tend to be more massive, and SNe that occur in less massive galaxies may be missed. As a result, there may be some bias toward sampling older stellar populations, since more massive galaxies tend to contain older, more metal-rich stars. However, we have also observed SNe 91bg hosts that were not targeted by BSNIP and in which 91bg-like SN occured more recently, in particular the host of SN2012fx, a nearby spiral galaxy for which the apparent SN explosion site is in the galactic bulge. The stellar population local to the SN explosion site is found to be relatively typical of those found in the BSNIP sample and has a weighted average stellar population age of $\sim 10\,\mathrm{Gyr}$.\\
We also compare our stellar population average ages and metallicities with those derived for the host galaxies of SNe Ia as observed by SDSS, based on a sample from \cite{Maoz2012}. There may be some bias associated with our measurement of stellar population ages for these SNe Ia host galaxies as the SDSS fiber is placed over the center of the galaxies they observe. This means that one is likely to observe most of the integrated light from the central, bulge regions of these galaxies, biasing our results to older ages as the star-forming, younger outer regions of some galaxies observed by SDSS are missed by the fiber. Consequently the weighted average plotted in figure \ref{fig:map} may not represent the exact properties of the stellar population close to the explosion site and therefore may represent an upper limit on the stellar population age.\\
We find that SNe Ia are hosted by galaxies that exhibit a broad range of weighted average stellar population ages, from the very young at $30\,\mathrm{Myr}$ to the very old ($\sim 10\,\mathrm{Gyr}$). This is in contrast to the stellar populations that host SNe 91bg, which cluster at the substantially older end of this distribution. This demonstrates that SNe 91bg may be associated with progenitors that tend to occur at longer delay times than SNe Ia (e.g. as suggested in \cite{Crocker17}). We find that while around half of our observed galaxies have H$_\alpha$ emission and nebular emission indicating recent star-formation activity, the majority of galaxies are dominated by absorption lines and have no strong emission lines. Moreover, with the exception of the host of SN1993aa, the regions of the galaxies that hosted 91bg-like SNe that exhibit emission lines are also found to contain older, $\sim10\,\mathrm{Gyr}$ stellar populations.\\
Note that while we find that 91bg-like SNe are associated with older populations than SNe Ia, this result does not preclude that the same galaxy could host both types of supernovae. Indeed, two examples of galaxies hosting both supernovae are SN1991bg and SN1957B (a SN Ia), which both occurred in M84, and of the 91bg-like supernova SN2006mr in Fornax A (\citep{Maoz08}, which also hosted five SNe Ia. The point here is that while the delay time distribution of SNe Ia may peak at 1 Gyr \citep{Childress2014}, SNe Ia can still occur in stellar populations with much greater ages (see Fig 3). Moreover,  both M84 and Fornax A are early-type galaxies, and will likely be dominated (in terms of number) by an old stellar population from which the 91bg-like supernova progenitors may arise but they may also contain younger populations that could host SNIa progenitors. It is important to note that while we find that 91bg-like SNe are associated with older populations than SNe Ia, our results are not contradicted by these observations. Our comparison sample of SNe Ia from \cite{Maoz2012} shows evidence that while the delay time distribution of SNe Ia may peak at $1\,\mathrm{Gyr}$ \citep{Childress2014}, SNe Ia can still occur in stellar populations with much greater ages (see Fig \ref{fig:map}). Moreover, M84 and Fornax A are early-type galaxies, and will likely be dominated by an old stellar population from which the 91bg-like supernova progenitors may arise. Thus it is possible for SNe Ia and 91bg-like supernovae to occur in the same host galaxy even if the delay time distribution of the progenitors is quite different.\\ 
We note that the MILES stellar population models we used with \texttt{pPXF} assumes the evolution of single stellar populations only. While tracking the stellar evolution and corresponding spectral morphology of single stars is relatively straightforward, it is fairly complicated in the case of interacting stars. Though the majority of stars in the Universe are likely part of a binary (or higher order) star system \citep{moe2017}, most spectral synthesis codes ignore the effects that binary (and multiple) stellar evolution have on the integrated light from galaxies due to the added layer of complexity (e.g. mass transfer and variation from single star evolutionary pathways) that must be taken into account. However, recent work has shown that incorporating binary star evolution models into spectral synthesis analysis codes does indeed result in different derived physical properties \citep{Stanaway2016}. On the other hand, the difference between single star spectral synthesis codes, such as GALEXEV, and binary spectral synthesis programs, like BPASS, is more prominent at early times, in particular for stellar ages $\lesssim 1$ Gyr \citep[see][fig. 2, right column]{Stanway2018}. Since the explosion sites that we are probing mostly consist of relatively old stellar populations, we are thus not overly concerned that our derived stellar properties are very different from the properties we would expect those stellar populations to have. A comparison of derived stellar properties with a spectral synthesis program that incorporates binaries (such as BPASS) is beyond the scope of this work, though based on the work of \citet{Stanway2018}, we would plausibly expect our derived stellar ages to decrease slightly, while the derived metallicities may slightly increase. 
\section{Summary and Conclusions}\label{sec:conclusion}
We obtained integral field observations of the apparent explosion sites of 17 91bg-like SNe. 11 of these observations yielded spectra of sufficiently high signal-to-noise to derive information about the stellar population within $\sim1\,\mathrm{kpc}$ projected radius of the explosion site. Using full-spectrum fitting, we found that the majority of the stellar populations that host 91bg-like SNe we observe are dominated by old, metal-rich stars. The majority lack evidence of recent star formation. We conclude that the SNe 91bg progenitors are unlikely to be associated with recent star formation, and likely have delay times $>6\,\mathrm{Gyr}$. This favors a progenitor model such as that proposed in \cite{Pakmor2013} or \cite{Crocker17}, which is composed of a He WD and CO WD.

\section*{Acknowledgements}
FHP is supported by an Australian Government Research Training Program (RTP) Scholarship. Parts of this research were conducted by the Australian Research Council Centre of Excellence for All-sky Astrophysics (CAASTRO), through project number CE110001020. IRS is supported by the Australian Research Council grant FT160100028. AJR is supported by the Australian Research Council grant FT170100243. BG is supported by the Australian Research Council grant FT140101202. The authors thank Rob Sharp, Fang Yuan, Stuart Sim, R\"{u}diger Pakmor, Friedrich R\"{o}pke, Anais Moller and Brian Schmidt for useful discussions, and the referee for their insightful comments to improve the manuscript. FHP thanks Peter Verwayen, Ian Adams and Henry Zovaro for their assistance with an occasionally temperamental telescope. 
This research has made use of the NASA/IPAC Extragalactic Database (NED), which is operated by the Jet Propulsion Laboratory, California Institute of Technology, under contract with the National Aeronautics and Space Administration.
This research has made use of the Digitized Sky Survey (DSS-2) based on photographic data obtained using The UK Schmidt Telescope. The UK Schmidt Telescope was operated by the Royal Observatory Edinburgh, with funding from the UK Science and Engineering Research Council, until 1988 June, and thereafter by the Anglo-Australian Observatory. The Digitized Sky Survey was produced at the Space Telescope Science Institute under US Government grant NAG W-2166.
We acknowledge the traditional owners of the land on which the ANU 2.3m telescope stands, the Gamilaraay people, and pay our respects to elders past and present.



\bibliographystyle{pasa-mnras}
\bibliography{PantherBib}

\end{document}